\documentclass[superscriptaddress,showpacs,aps,prl,twocolumn]{revtex4}
\usepackage{graphicx}
\usepackage{amsmath}
\usepackage[english]{babel}
\usepackage{psfrag}
\usepackage[latin1]{inputenc}

\newcommand{\lpar}{lin\,{\parallel}\,lin} 
\newcommand{\lperp}{lin\,{\perp}\,lin}      
\newcommand{\hpar}{h\,{\parallel}\,h}     
\newcommand{\hperp}{h\,{\perp}\,h}         

\newcommand{\LOD}{Laboratoire Ondes et D\'{e}sordre, FRE 2302 du CNRS,
1361 route des Lucioles, 06560 Valbonne, France}
\newcommand{\LKB}{Laboratoire Kastler Brossel, Universit\'e Pierre et Marie Curie,
Case 74, 4 place Jussieu, 75252 Paris, France}
\newcommand{\MPIPKS}{Max-Planck-Institut f\"ur Physik komplexer
Systeme, N\"othnitzer Str.\ 38, D-01187 Dresden, Germany}

\begin{document}
\title{Coherent backscattering of light by cold atoms:   
theory meets experiment}
\author{Guillaume Labeyrie}
\affiliation{\LOD}

\author{Dominique Delande}
\affiliation{\LKB}

\author{Cord A. M\"uller}
\affiliation{\LOD} \affiliation{\MPIPKS}

\author{Christian Miniatura}
\affiliation{\LOD}

\author{Robin Kaiser}
\affiliation{\LOD}

\date{\today}

\begin{abstract}
Coherent backscattering (CBS) of quasi-resonant light by cold atoms 
presents some specific
features due to the internal structure of the atomic scatterers.
We present the first quantitative  
comparison between the experimentally
observed CBS cones and Monte-Carlo calculations which take into
account the shape of the atomic cloud as well as the internal atomic
structure. 
\end{abstract}

        
\maketitle

When 
light is elastically scattered off an optically thick disordered medium,
the interference between all partial waves 
produces strong angular fluctuations of the intensity distribution 
(speckle pattern).
Averaging over the positions of the scatterers 
washes out interferences and produces a smooth 
reflected diffuse intensity, \textit{except} around backscattering where 
it is enhanced. This 
coherent backscattering effect (CBS), originates from a two-wave 
interference, namely 
the interference between waves traveling along the same scattering paths 
but in reverse order~\cite{sheng,wavediffusion}. This interference 
is constructive 
at exact backscattering and only survives, after averaging, in a narrow 
angular range 
$\Delta\theta \simeq 1/k\ell$ around it 
($k$ is the light wave-number and $\ell$ the scattering mean free path). 
The CBS enhancement factor measures the ratio of the maximum intensity 
(measured at backscattering) to the background intensity measured at 
angles $\theta \gg 1/k\ell$. 
Because it originates from a two-wave 
interference, the enhancement factor is
bounded by 2. 

For vector waves like light,  
either linear or circular polarization can be considered. 
This leads to four different
polarization channels in scattering experiments: 
$\lpar$ where both the incoming and outgoing photons
are linearly polarized along the same axes, 
$\lperp$ where they are
linearly polarized along orthogonal axes, $\hpar$ where they are both
circularly polarized with the same helicity 
(because they propagate in opposite directions, they
have opposite polarizations) and $\hperp$ where they are
circularly polarized with opposite helicities, {\em i.e.} same polarization.

For spherically-symmetric scatterers (and hence 
for point-dipole scatterers) and in the $\hpar$ channel 
the CBS enhancement factor
is equal to 2~\cite{Wiersma95} because the following
conditions are met: 
$(i)$ All scattering paths have a distinct reverse counterpart;
$(ii)$ The two paths in each pair contribute with the same amplitude and
phase, so that a maximum interference contrast is guaranteed.
Condition $(i)$ is not true for
single scattering paths which thus do not contribute to CBS.
In general, this implies an enhancement factor slightly 
smaller than 2. However,
for spherically-symmetric scatterers, there is no single scattering 
signal in the backward direction 
in the $\lperp$ and $\hpar$ channels.
Condition $(ii)$ is met in the
parallel channels $\hpar$ and $\lpar$ as a general consequence of reciprocity 
\cite{reciprocity}
(which is equivalent to time-reversal symmetry in the absence of absorption). 
Therefore, an enhancement factor much smaller than 2 has only been observed in
the perpendicular channels or by breaking the reciprocity with the help of an
external magnetic field~\cite{cbswithB}. 

The recent experimental observation of CBS of
light by an optically thick sample of
laser-cooled Rubidium atoms~\cite{Labeyrie1,Labeyrie2} 
reported -- in all channels --
surprisingly small enhancement factors, 
typically between 1.05 and 1.2.   
A qualitative explanation of the experimental results
in the double scattering picture 
has been given in~\cite{Labeyrie2,Jonckheere,details}.  
The Rubidium atoms, being much smaller than the optical 
wavelength, are point-scatterers, but do not behave like classical dipoles: 
the atomic Zeeman internal structure leads to a violation of the 
previous conditions
for a perfect CBS enhancement.   
Condition $(i)$ is not met because single scattering is present
in all polarization channels, due to transitions between different atomic
sub-states. 
Condition $(ii)$ is not met because the amplitudes associated with  
the direct and reverse scattering path 
are in general {\em not equal}.
This imbalance of the amplitudes reduces the interference contrast, and leads
to a CBS cone much smaller than for point-dipole scatterers.

In this paper, we present 
the first extensive comparison between the 
experimentally measured enhancement
factors and CBS
cone shapes (in all four polarization channels) 
and a theoretical calculation
which takes into account the two most important ingredients of
the experiment: the atomic internal structure and the
peculiar shape of the scattering medium, an approximately spherical
cloud of cold atoms produced in a magneto-optical trap (MOT).
 
The experimental setup is described in detail in~\cite{Labeyrie2}.
The probe laser beam is tuned on the $D2$ line 
$5S_{1/2} \to 5P_{3/2}$ of Rb$^{85}$ at $\lambda = 780$ nm
 and is resonant 
with the corresponding hyperfine ($F=3 \to F'=4$) 
transition (natural line-width $\Gamma /2\pi =5.9$ MHz, on-resonant 
light scattering cross-section $\sigma = \frac{(2F'+1)}{3(2F+1)} 
\frac{3\lambda^{2}}{2\pi}$). 
An atomic cloud, containing about  ${\cal N}=7\times 10^{9}$ atoms, 
is produced in a MOT at a temperature in the 100 $\mu $K
range (residual {\it rms}-velocity spread about 10 cm.s$^{-1}$). 
Imaging techniques show that atoms in the cloud are 
distributed with a quasi-Gaussian density of FWHM ($x$,$y$,$z$)-dimensions 
5.88 mm$\times $4.87 mm$\times $4.63 mm (axis $y$ is the polarization axis 
of the CBS probe light in the linear channels and axis $z$ its 
propagation axis). 
The on-resonance mean free path at the center of the sample 
$\ell = 1/(n_0\sigma)$ is of the order
of 200 $\mu$m, much larger than the wavelength ($n_0$ is the atomic density at 
the trap center).
We are thus in the dilute regime where $k\ell \gg 1$. 
The typical width of the CBS cone 
is of the order of 0.3 to 0.9 mrad, sufficiently above the resolution 
limit of the apparatus (0.1 mrad).

The numerical calculation of the CBS cone
takes into account the internal structure of the atom.
Since all possible hyperfine dipole transitions for the $D2$ line are 
well separated 
on the scale of the line-width $\Gamma$, 
we only consider quasi-resonant light 
scattering induced by the above-mentioned {\em closed} hyperfine transition. 
We indeed expect it to give the dominant contribution to scattering 
(the nearest transition $F=3 \to F'=3$ lies 20 $\Gamma$ away). Note 
however that the contribution of other transitions is currently under 
investigation \cite{Havey}. When an atom scatters the incoming light, it
may stay in the same Zeeman sub-level --- this is a Rayleigh transition ---
or change its magnetic quantum number --- this is a degenerate Raman transition.
In both cases, at weak laser intensities and if recoil and Doppler 
effects are negligible
(which one expects to be the case for our cold atomic cloud), 
the scattered photon 
has the same frequency than the incoming photon: the scattering is 
{\em elastic}.
The scattering amplitude by a single atom 
 depends on the initial and final Zeeman sub-levels, 
 on the scattering direction
 and on the incident and scattered polarizations~\cite{details}. 
As the atoms produced
 in a MOT are not in well-defined internal states, but rather in
 a statistical mixture of Zeeman states, the calculation of the
 CBS cone requires, in addition to the usual position 
averaging, an averaging over the possible internal ground-states of the atom.
We choose to perform the average over the positions of the
scatterers with a Monte-Carlo method and to use an internal
analytical average by employing the average atomic
scattering vertex~\cite{details,allorders}. 
The details of the
method are given in~\cite{bouleg}. The essential assumption 
is that all Zeeman 
sub-levels in the ground-state are equally populated, without
any coherence between sub-levels. This is a reasonable assumption
provided no optical pumping takes place in the medium
(see discussion below).
Our Monte-Carlo  method is flexible, as it makes it possible to 
compute the CBS cone
for an arbitrary spatial 
repartition of the scatterers. Furthermore, it allows us to take into account 
some rather small, yet not negligible effects, such as the 
non-uniform incoming intensity sent on the sample 
(because a Gaussian laser beam and diaphragms are used). 
It is important to note that all the parameters entering the
numerical calculations (optical thickness of the medium,
shape and dimensions of the atomic cloud, geometrical properties
of the incoming laser beam) have been experimentally
{\em measured} which means that the CBS cones that we calculate
have {\em no} adjustable parameter.

\begin{figure}
\psfrag{hperph channel}{$\hperp$}
\psfrag{hparh channel}{$\hpar$}
\psfrag{linparlin channel}{$\lpar$}
\psfrag{linperplin channel}{$\lperp$}
\centering \includegraphics[width=0.85\textwidth,angle=-90]{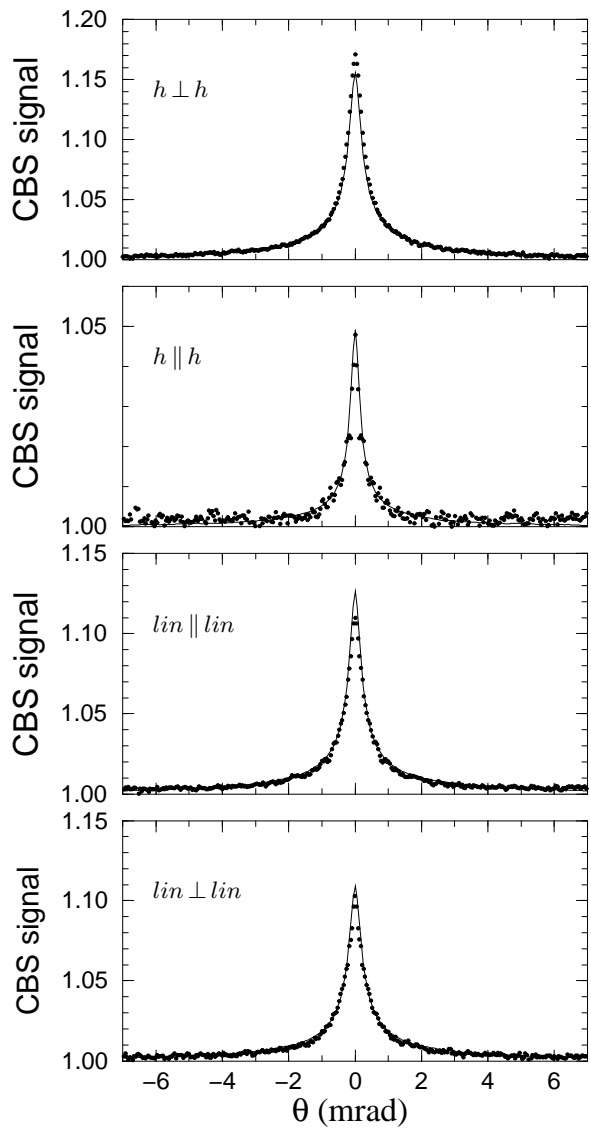}
\caption{The CBS cones experimentally observed on a quasi-spherical
cloud of cold Rubidium atoms (optical thickness 26) illuminated
by a weak resonant laser beam, in the four polarization
channels. 
The CBS cones have been angularly averaged in order to improve
the signal/noise ratio. The solid lines are the theoretical
predictions of a Monte-Carlo calculation taking into account
the internal atomic structure, the shape of the scattering medium,
and the geometry of the incoming laser beam. All the parameters
are experimentally {\em measured}, so that there is {\em no}
adjustable parameter. The shapes and angular widths of the various
cones are very well reproduced, including the wings of
the cones and details like the small angular width in the $\hpar$ 
channel.}
\label{4cones_at}
\end{figure}

Figure~\ref{4cones_at} 
shows the CBS cones recorded for an atomic cloud 
of optical thickness $b=\sqrt{2\pi} r_0/\ell =$ 26 in the four
polarization channels, compared with the cones computed using our Monte-Carlo
method ($r_0$ is the {\em rms}-radius of the atomic cloud along the $z$-axis). 
The plotted cones are an angular average of 
the 2D images recorded on the CCD. This data processing improves 
the signal-to-noise ratio at the expense of smearing out 
any angular anisotropies, either due to 
the scattering itself or to the shape 
of the scattering medium. In the circular polarization channels, 
this proves to be safe since the recorded cones are isotropic (although 
the scattering medium itself has not a perfect cylindrical symmetry around the
probe propagation axis).
In the $\lpar$ channel, the cone shape is anisotropic (roughly an elliptical
shape, see table~\ref{table}) and the widths
are different along the direction of the incoming polarization and 
perpendicular
to it~\cite{Rb:B}. In the $\lperp$ channel, the CBS cone has a 
four-fold symmetry~\cite{Rb:B}
and two different widths can be measured, either along one of 
the polarization axes
or at $45^{\circ}$ from it. 
The FWHM angular widths 
have been measured in the various configurations and are
displayed in table~\ref{table}, together with the
result of the Monte-Carlo calculation. 

For the 
enhancement factor, the most dramatic effect of the atomic internal 
structure, already noticed
in the first experimental results~\cite{Labeyrie1,Labeyrie2} is that the
``best" channel for point-dipole scatterers ($\hpar$)  
is the worst one 
for the atomic scatterers. Conversely, the worst channel for point-dipole
scatterers ($\hperp$) is the best one for atomic scatterers. 
This has been already established in~\cite{Jonckheere}
and further justified
in \cite{details} by carefully  
analyzing the atomic scattering vertex.
But as a consequence of the
approximations done (analysis limited to single and double scattering 
in a semi-infinite medium with constant atomic density), the CBS cone 
computed in~\cite{Jonckheere} had only roughly the correct angular
width (and the mean free path had to be adjusted) while its shape was
not in excellent agreement with the experimental result,
especially in the wings. Moreover, the enhancement factors
were only fairly well reproduced
by the calculation. 

The present agreement 
is much more satisfactory because it 
takes into account both higher-order scattering and the
geometrical effects induced by the peculiar shape of the
medium.
The cone shapes are very well reproduced as well as the angular
widths. Note that the wings of the cones
are well reproduced and that non-trivial details such
as the fact that the CBS cone is significantly narrower in
the $\hpar$ channel than in the $\hperp$ one are also well
predicted by our calculation. 
This is a strong evidence that
our theoretical approach catches the most important aspects of
multiple scattering of light by cold atoms. It also indicates
that most of the experimental parameters are under control.

However, there are 
small differences for the angular widths,  
at most $0.15$ mrad, 
which is a statistically significant deviation. Complementary Monte-Carlo
calculations, discussed in~\cite{bouleg}, show that the angular
width is rather sensitive to the detailed shape of the medium, especially
in the external layers of the atomic cloud. Keeping the same
optical thickness, but changing the density of the medium from
a Gaussian density $\exp(-r^2/2r_0^2)$ to a $\exp(-r^4/4r_0^4)$
density (i.e. sharper edges) increases the angular width by
more than 50\%, the precise value depending on the polarization channel.
As the atomic density in a MOT is not precisely a Gaussian function,
we attribute the small differences in the angular widths to 
an imperfect control of the shape of the medium. 
 
Another noticeable difference is that
the enhancement factor is slightly
underestimated in the $\hperp$ 
channel and slightly overestimated
in the linear channels.

\begin{table}
\begin{tabular}{c|cccc} 
Channel         & $\alpha$(exp)  &   $\alpha$(th) & $\Delta\theta$(exp) & $\Delta\theta$(th)\\ 
     \hline
$\hperp$                &  1.171  & 1.156 & 0.61 & 0.58\\  
$\hpar$                 &  1.048  & 1.049 & 0.30 & 0.40\\  
$\lperp$ \mbox{averaged}&  1.103  & 1.109 & 0.62 & 0.58\\  
$\lpar$  \mbox{averaged}&  1.110  & 1.126 & 0.62 & 0.54\\ 
$\lpar$  \mbox{scanpar }&    "    &  "     & 0.91 & 0.76\\ 
$\lpar$  \mbox{scanperp}&    "    &  "     & 0.51 & 0.44\\ 
     \hline
\end{tabular}
\caption{{\em Enhancement factor $\alpha$ and angular width $\Delta\theta$
(FWHM in mrad) experimentally observed in the various polarization
channels (exp), compared to the results of 
our Monte-Carlo calculation (th) without {\em any}
adjustable parameter. 
``scanpar'' and ``scanperp'' refer to the widths measured in the 
direction parallel to
the incoming polarization, resp.\  perpendicular to it. 
The experimental uncertainty on the widths is $\pm 0.03$
mrad, while the enhancement factors are measured $\pm 0.006$ 
($\pm $2 standard deviations).} 
}
\label{table}
\end{table}

\begin{figure}
\centering \includegraphics[width=0.4\textwidth,angle=-90]{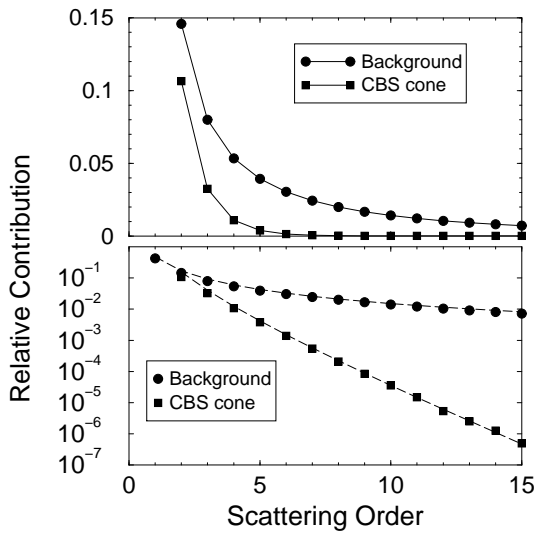}
\caption{Contributions of the successive multiple scattering orders 
to the smooth background and to the CBS cone peak value, in the 
experimental conditions of fig.~\protect\ref{4cones_at} in the
$\hperp$ channel. All contributions are
normalized such that the total background is unity. The single scattering
contribution (background only) is the strongest one, 42.6\% of the total.
In the upper plot (linear scale), one can see the slow decrease
of the contributions to the background, which shows that we are indeed
in the multiple scattering regime. 
The CBS contributions decrease very rapidly, which indicates that
low orders of scattering are the essential contributions. On a logarithmic
scale (lower plot), one clearly sees that the contributions to the
background decrease algebraically with the scattering order $N$,
roughly following the prediction of the diffusion approximation
$\propto N^{-3/2}$ (dashed line). This proves that the finite geometry,
which cuts the very long scattering paths, is not crucial here.
The CBS contributions decrease exponentially with $N,$ in excellent agreement
with the analytic prediction using the properties of the
atomic scattering vertex~\protect\cite{TheseCord} 
$\propto (19/40)^N \times N^{-3/2}$,
shown as a dashed line.}
\label{order3}
\end{figure}

In fig.~\ref{order3}, we show, for the parameters of fig.~\ref{4cones_at}, 
the contributions of the various
orders of scattering
to the background and the CBS cone peak value. The quantities are easily
extracted from the Monte-Carlo calculation. One can make several observations.
Firstly, the background contributions decrease rather slowly with the
scattering order $N.$ For a semi-infinite medium, the standard diffusion 
approximation predicts that it decreases like $N^{-3/2}$ in excellent agreement
with our numerical observation~\cite{diffusion}.  
At very large order, the finite
size of the medium implies a faster decrease (very long scattering paths
unavoidably escape the medium), but this effect in negligible for orders
lower than few tens. On the contrary, the contributions to the CBS cone
decrease {\em exponentially} with $N.$ In the case of the
$F=3\to F'=4$ transition, theory~\cite{TheseCord,Akkermans02} 
predicts a decay like $N^{-3/2} \times (19/40)^N$ in excellent 
agreement with our
numerical results.  This unambiguously proves that, for atoms 
with a degenerate ground-state, the CBS effect is mainly dominated by 
low-order scattering and the properties of the atomic scattering vertex.
This also explains why the CBS cone computed in~\cite{Jonckheere}
gives a reasonable estimate of the enhancement factor.
Indeed, single and double scattering contribute to roughly
57\% of the background and 68\% of the CBS cone. Forgetting higher
orders is thus a rather good approximation, even if the medium
is optically thick. 

\begin{figure}
\psfrag{hpar}{$\hpar$}
\psfrag{hperp}{$\hperp$}
\psfrag{linparlin}{\small{$\lpar$}}
\psfrag{linperplin }{\small{$\lperp$}}
\centering \includegraphics[width=0.32\textwidth,angle=-90]{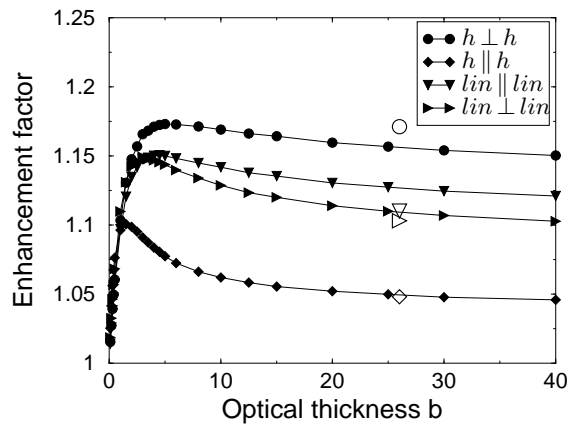}
\caption{Enhancement factor of the CBS cone
as a function of the optical thickness
of the atomic cloud, computed for the experimental parameters,
in the various polarization channels (the
points at optical thickness 26, corresponding to data
in fig.~\protect\ref{4cones_at} and table~\protect\ref{table}
are not shown for clarity). The experimental points
(open symbols)
are in good agreement (relative error in the heights
smaller than 14\%), with the $\hpar$ channel giving the largest cone,
followed by the $\lpar$ and $\lperp$ channels, the $\hpar$ cone 
being significantly
smaller. }
\label{ef_at_bg}
\end{figure}

In order to understand the small discrepancy in the enhancement
factors,
we calculate the enhancement factor as a function of the optical thickness,
in the four polarization channels, using the experimentally measured
parameters. The result is shown in fig.~\ref{ef_at_bg}, 
 together with the
four experimental points measured at optical thickness $b=26.$
It is clear that in this regime the computed enhancement factor 
depends only weakly 
on the optical thickness. This is because the CBS cone
is mainly due to low order multiple scattering, which takes
place in the external layers
of the atomic cloud (on the side of the incoming laser beam), 
and is thus only weakly sensitive to the deep
layers and the optical thickness. Complementary calculations~\cite{bouleg}
also prove that the enhancement factor 
-- contrary to the angular width -- is not very sensitive to the detailed
shape of the medium.

The relative strength of the four CBS cones is well reproduced by the 
calculation: the $\hperp$ cone is the most important one, 
followed by the $\lpar$ then $\lperp$ cones, the $\hpar$ one being
significantly smaller. The deviation in the enhancement factor
between the calculation and the observation is less than 0.02,  
{\em i.e.} the  height
of the cone is reproduced with a relative error smaller than 14\%.
 
We are not completely sure of the reason of such a small discrepancy. 
We are inclined to attribute it to a non-uniform
distribution of the atomic state over the various Zeeman sub-levels. 
Even after the trapping beams and the magnetic field of the MOT are switched
off, it may remain some polarization of the atomic ground state.
Another possibility -- more likely in our opinion -- is that
the CBS probe beam, although weak, induces some optical pumping.
This effect is difficult to estimate
in optically thick media, because
the atoms are exposed not only to the incoming beam but also to the light
scattered by other atoms. The importance of optical pumping seems to be 
supported by preliminary 
experimental results which suggest
that the discrepancy between the calculation and
the experiment is reduced when the number of exchanged photons is 
decreased.

To summarize, we have presented in this paper
the first 
comparison between experimentally observed 
CBS cones on a could of cold atoms  
and  
Monte-Carlo simulations including the atomic internal structure and the 
peculiar sample geometry, without any
adjustable parameter. 
The agreement is very good for the cone shapes
and angular widths, as well as for the enhancement factor.

We thank CNRS, the PACA Region and the  Groupe de Recherche 
PRIMA for financial support.
Laboratoire Kastler-Brossel de l'Universit\'e Pierre et Marie Curie et
de l'Ecole Normale Sup\'erieure is UMR 8552 du CNRS. CPU on a Cray SX5 computer
has been provided by IDRIS.


\begin{thebibliography}{99}

\bibitem{sheng} P. Sheng, {\em Introduction to Wave Scattering,
Localization and Mesoscopic Phenomena}, Academic Press (1995).

\bibitem{wavediffusion} {\em New Aspects of Electromagnetic
and Acoustic Wave Diffusion}, edited by POAN Research Group, 
Springer Tracts in Modern Physics, vol. \textbf{144}, Springer, 
Berlin (1998).

\bibitem{Wiersma95} Wiersma D.S., van Albada M.P., van Tiggelen B.A. and 
Lagendijk A., {\em Phys.\ Rev.\ Lett.}, \textbf{74} (1995), 4193.

\bibitem{reciprocity} van Tiggelen B.A. and Maynard R., in 
\textit{Wave Propagation in Complex media}, IMA Vol. 96, edited by 
G. Papanicolaou (Springer, New York, 1997), 252.

\bibitem{cbswithB} Lenke R. and Maret G., {\em Eur. Phys. J. B}, 
\textbf{17} (2000), 171. 

\bibitem{Labeyrie1} Labeyrie G., de Tomasi F., Bernard J.-C., 
M\"{u}ller C.A., Miniatura C. and Kaiser R., {\em Phys.\ Rev.\ Lett.}, 
\textbf{83} (1999), 5266.

\bibitem{Labeyrie2}  Labeyrie G., M\"{u}ller C.A., Wiersma D.S., 
Miniatura Ch. and Kaiser R., {\em J. Opt. B: Quantum Semiclass. Opt.}, 
\textbf{2} (2000), 672.

\bibitem{Jonckheere}
Jonckheere T., M\"uller C.A., Kaiser R., Miniatura C. and
Delande D., {\em Phys.\ Rev.\ Lett.}, \textbf{85} (2000), 4269.

\bibitem{details} M\"uller C.A., Jonckheere T., Miniatura C. and
Delande D., {\em Phys.\ Rev. A}, \textbf{64} (2001), 053804.

\bibitem{Havey} M. Havey, private communication.

\bibitem{allorders} M\"uller C.A. and Miniatura C., 
{\em J. Phys. A: Math. Gen.} \textbf{35} (2002), 10163, \url{physics/0205029}. 

\bibitem{bouleg} Labeyrie G., Delande D., M\"uller C.A., Miniatura C. 
and Kaiser R., submitted to {\em Phys.\ Rev. A} (2002). 

\bibitem{Rb:B} Labeyrie G., Miniatura Ch., M\"uller C.A., Sigwarth O., 
Delande D. and Kaiser R., {\em Phys. Rev. Lett.}, \textbf{89} (2002), 
163901.

\bibitem{diffusion} van der Mark M.B., van Albada M.P. and Lagendijk A., 
{\em Phys.\ Rev. B}, \textbf{37} (1988), 3575. 

\bibitem{Akkermans02} Akkermans E., M\"uller C.A. and Miniatura Ch., 
submitted to {\em Phys. Rev. Lett.} (2002), \url{cond-mat/0206298}.

\bibitem{TheseCord} M\"uller C.A., PhD thesis (Universities of 
Munich/Nice-Sophia Antipolis, 2001),  
\url{http://www.ub.uni-muenchen.de/elektronische_dissertationen/physik/Mueller_Cord.pdf}.

\end{thebibliography}
\end{document}